\documentclass[prlb,aps,twocolumn,
floats,superscriptaddress]{revtex4-1}
\usepackage{xcolor}
\usepackage{amsmath}
\usepackage{graphicx}
\usepackage{amssymb}
\usepackage{epsfig}
\usepackage[colorlinks = true]{hyperref}

\newlength{\upit}\upit=0.1truein

\newcommand{\ltappr}{{{\lower4pt\hbox{$<$} } \atop \widetilde{ \ \ \ }}}
\newlength{\bxwidth}\bxwidth=1.5 truein

\newcommand{\tr}{{\hbox{Tr}}}

\newcommand{\dg}{^{\dagger }}

\newcommand{\gtappr}{{{\lower4pt\hbox{$>$} } \atop \widetilde{ \ \ \ }}}

\newcommand{\up}{\uparrow}
\newcommand{\dw}{\downarrow}

\newcommand{\bk}{{\bf{k}}}
\newcommand{\bx}{{\bf{x}}}

\newcommand{\ua}{\uparrow}
\newcommand{\da}{\downarrow}

\newcommand{\be}{\begin{equation}}
\newcommand{\ben}{\begin{equation*}}
\newcommand{\ee}{\end{equation}}
\newcommand{\een}{\end{equation*}}

\newcommand{\bmx}{\begin{array}}
\newcommand{\emx}{\end{array}}
\newcommand{\bean}{\begin{eqnarray*}}
\newcommand{\eean}{\end{eqnarray*}}

\newcommand{\pref}[1]{(\ref{#1})}

\newcommand{\bra}[1]{\left\langle #1 \right\vert}
\newcommand{\ket}[1]{\left\vert #1\right\rangle}
\newcommand{\braket}[1]{\left\langle #1\right\rangle}
\newcommand{\mat}[1]{\left(\bmx{cc}#1\emx\right)}
\newcommand{\matc}[2]{\left(\bmx{#1}#2\emx\right)}

\newcommand{\red}[1]{{#1}}

\newcommand\ltdash{\raise-0.7pt\hbox{$\scriptscriptstyle |$}}

\newlength{\figwidth}
\newlength{\shift}
\begin{document}
\title{The Triplet Resonating Valence Bond State and Superconductivity 
in Hund's Metals
}
\author{Piers Coleman}
\affiliation{
Center for Materials Theory, Department of Physics and Astronomy,
Rutgers University, 136 Frelinghuysen Rd., Piscataway, NJ 08854-8019, USA}
\affiliation{Department of Physics, Royal Holloway, University
of London, Egham, Surrey TW20 0EX, UK.}
\author{Yashar Komijani}
\affiliation{
Center for Materials Theory, Department of Physics and Astronomy,
Rutgers University, 136 Frelinghuysen Rd., Piscataway, NJ 08854-8019, USA}
\author{Elio J. K\"onig}
\affiliation{
Center for Materials Theory, Department of Physics and Astronomy,
Rutgers University, 136 Frelinghuysen Rd., Piscataway, NJ 08854-8019, USA}
\date{\today}
\pacs{PACS TODO}
\begin{abstract}
{ 
A central idea in strongly correlated systems is that doping a Mott insulator leads to a superconductor by transforming the resonating valence
bonds (RVBs)  into spin-singlet Cooper pairs.  Here, we argue that a 
spin-triplet RVB (tRVB) state, driven 
by spatially, or orbitally  anisotropic ferromagnetic interactions
can provide the parent state for triplet superconductivity. 
We apply this idea
to the iron-based superconductors, arguing that 
strong onsite Hund's interactions develop  intra-atomic tRVBs between the
t$_{2g}$ orbitals. 
On doping, the presence of two iron atoms per unit cell allows these
inter-orbital triplets to coherently delocalize onto the Fermi surface,
forming  a fully gapped triplet superconductor. This mechanism
gives rise to a unique 
staggered structure of onsite pair correlations, detectable
as an alternating $\pi$ phase shift in a scanning tunnelling Josephson microscope.
}
\end{abstract}

\maketitle
Thirty years ago, 
Anderson proposed\,\cite{Anderson1987} the intriguing idea that the 
resonating valence bonds (RVBs) of a spin liquid could, on doping,
provide the fabric for the development of unconventional
superconductivity.   A key aspect of the RVB theory, is that it 
departs from weak-coupling approaches to superconductivity,
positing that instead of a pairing glue,  
superconductivity develops from the 
entangled pairs already present in a spin liquid. 
RVB theory provides a
natural account of the connection between  d-wave pairing and
antiferromagnetism\,\cite{Lee06} in almost-localized systems, 
a connection that has proven invaluable to
the understanding of many families of superconductors, from the
cuprate superconductors, to their miniature cousins, the 
115 heavy-fermion compounds \cite{Thompson2011}. 

However, to date, there is no counter-part of RVB theory that applies
to ferromagnetically correlated  materials. 
There are 
a wide variety of unconventional
superconductors which, to some extent or another, involve
strong ferromagnetic (FM) spin correlations.  
Examples include
uranium-based heavy fermion materials~\cite{Joynt02,Pfleiderer2009} 
which lie close to a
FM quantum critical point, candidate low-dimensional triplet superconductors such as the Bechgaard salts~\cite{LangMueller2008}, twisted double bilayer graphene~\cite{LiuKim2019,ShenZhang2019},
and  various transition
metal superconductors \cite{Stewart11,Hosono2018}, notably the iron-based and ruthenate
superconductors, which as Hund's metals involve strong local
FM correlations between orbitals. Various papers have speculated that the Hund's interactions might provide the
origin of the pairing in these systems \cite{PuetterKee2012,HoshinoWerner2015,VafekChubukov2017,CheungAgterberg2019,AndersonTriplet,Norman94}.

Is there a
ferromagnetic analog to the RVB pairing mechanism?
Here we build on an 
observation~\cite{Cerge} that magnetic anisotropy 
in a ferromagnet plays an analogous role to frustration in an
antiferromagnet (AFM), generating 
a fluid of triplet resonating valence bonds (tRVBs). 
We propose  that like their singlet cousins, tRVB states 
can, on doping, lead to the development of
triplet pairing. One of the exciting features of this idea, is that 
tRVBs can form 
within the interior of Hund's coupled atoms, which 
under the right symmetry
conditions\,\cite{AndersonTriplet,Hotta04} can
coherently tunnel into the bulk to develop 
triplet superconductivity~\cite{Yin2011,Georges2013}. 
\begin{figure}
\includegraphics[width=\linewidth]{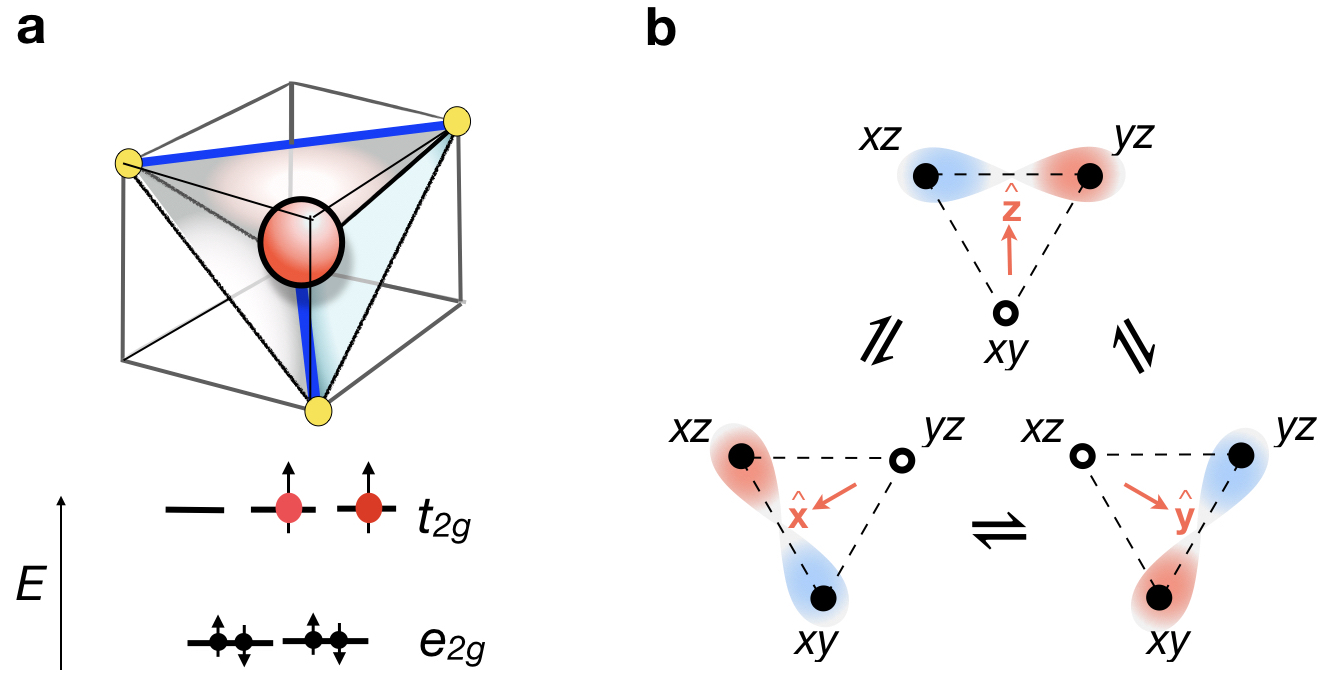}
\caption{\small a) Isolated tetrahedron in iron-based superconductors,
showing the two electrons forming a $S=1$ triplet in the t$_{2g}$
orbitals.
c) Triplet resonating valence bond {(tRVB) as the ground state of a
Hund's metal atom}. The blue and red colors reflect the odd parity of the triplet pairs, while the red arrows denote the quantization axis
(d-vector) of the $m=0$ triplet pair. 
}
\label{fig:tRVBIllust}
\end{figure}

Consider
an easy-plane FM interaction
$H_{ij}=-J\vec{S}_{i}\cdot \vec{S}_{j}+\Delta J
S^{z}_{i}S^{z}_{j}$, ($J>0$) between two spin-1/2 moments $\vec{ S}_{i}$ and
$\vec{S}_{j}$. 
In the Heisenberg limit ($\Delta J=0$) and in the presence of a small symmetry breaking Weiss field, the 
ground-state is
a product state which lacks entanglement. Suppose
the magnetization points in the $x$ direction,
the product ground-state can then be written in
terms of
triplets, 
\begin{equation}\label{}
\frac{\ket{\ua_i}+\ket{\da_i}}{\sqrt
2}
\frac{\ket{\ua_j}+\ket{\da_j}}{\sqrt
2}
=\frac{\ket{\ua_i\ua_j}+\ket{\da_i\da_j}}{2}+\frac{\ket{\ua_i\da_j}+\ket{\da_i\ua_j}}{2} .  
\end{equation}
An easy-plane anisotropy ($\Delta J>0$) 
projects out the equal-spin pairs on the right-hand-side, stabilizing 
an entangled spin-1 ground state with $m_{z}=0$. 
In the corresponding easy-plane ferromagnet, with Hamiltonian
$H= \sum_{(i,j)}H_{ij}$, the intersite
couplings preserve the $m_{z}=0 $ structure of the valence bonds,
and the resulting ground-state is a quantum superposition of triplet
pairs which retains its ferromagnetic correlations, and may even exhibit
long-range order.\cite{Supplement} 

Our interest in a tRVB ground-state lies in its potential as 
a pre-entangled parent state of a triplet superconductor. 
In classic RVB theory, an antiferromagnetic superexchange interaction,
is decoupled in terms of singlet pairs~\cite{KotliarLiu1988}:
\begin{equation}\label{2}
J \vec{S}_{i}\cdot \vec{S}_{j}\equiv 
-\frac{J}{2}
\bigl(
\psi \dg _{i\uparrow}\psi \dg _{j\dw}- \psi \dg _{i\dw}\psi \dg _{j\up}
\bigr)
\bigl(
\psi _{j\dw} \psi _{i\up}- \psi _{j\up}\psi  _{i\dw}\bigr), 
\end{equation}
where we have used a fermionic 
representation of the spins, 
$\vec{S}_{j} = \psi_{j}\dg \left(\frac{\vec{\sigma }}{2} \right)\psi _{j}$.
The corresponding relation for triplet valence bonds is obtained
by rotating the spin co-ordinate system at site $j$
through
180$^\circ$ about the z-axis, which gives
\begin{eqnarray}
&-&J_{A} ( S^{x}_{i}S^{x}_{j}+ S^{y}_{i}S^{y}_{j}-S^{z}_{i}S^{z}_{i})\cr
&&\equiv
-\frac{J_{A}}{2}
\bigl(
  \psi \dg _{i\uparrow}\psi \dg _{j\dw}+ \psi \dg _{i\dw}\psi \dg _{j\up}
\bigr)
\bigl(
\psi _{j\dw} \psi _{i\up}+ \psi_{j\up}\psi  _{i\dw}\bigr), 
 \label{eq:AnisotropicIA}
\end{eqnarray}
demonstrating how xy anisotropy stabilizes a triplet pair.

The most direct application of the tRVB idea 
considers an easy-plane Heisenberg ferromagnet: by analogy with the
singlet RVB pairing mechanism, doping with holes drives the formation
of a triplet superconductor.  
On a  square lattice,  this scenario leads to a p$_{x}$+ ip$_{y}$
triplet superconductor, to be presented elsewhere. A more dramatic possibility, 
in which $i$ and $j$ represent orbitals
of a single  atom,  permits us to
apply the tRVB idea to Hund's coupled metals. Here an application of
particular current interest, is as a theory for iron-based superconductors (FeSC). 

The family of FeSC are characterized by high
transition temperatures with a fully gapped Fermi surface. The
presence of antiferromagnetic correlations and a marked Knight shift
has led to the long-held assumption that these materials are spin
singlet superconductors \cite{Mazin2008,Stewart11}. On the other hand, the recent observation \cite{Lee18} of a robust ratio $2\Delta/T_c\sim 7.2$ between
the gap $\Delta$ and the transition temperature $T_c$ across \red{a
broad range} of FeSC 
motivates the search for a common pairing mechanism\red{, one that is robust
against the wide spectrum
of Fermi surface morphologies, and hence most likely rooted in
the local electronic structure of the iron atoms. 
Here,} 
we propose that these systems are
tRVB superconductors, with a fully gapped Fermi surface, an anisotropic Knight
shift and an alternating pair wave-function. 

The symmetry properties 
of a Hund's coupled triplet superconductor were 
first considered by Anderson\,\cite{AndersonTriplet}, 
who observed that in systems with a center of inversion, the
odd-parity wavefunction of a triplet condensate 
prevents onsite
triplet pairing unless the lattice has an even number of atoms per
unit cell, related to each other via inversion. 
In this situation, the odd-parity nature of the condensate means
that the onsite pair wavefunction reverses sign
when reflected through the center of inversion 
\begin{equation}\label{invert}
\langle \psi_{a\sigma } (\mathbf{x})\psi_{b\sigma '} (\mathbf{x})\rangle = 
-\langle \psi_{a\sigma } (-\mathbf{x})\psi_{b\sigma '} (-\mathbf{x})\rangle,
\end{equation}
where $a$ and $\sigma $ are the orbital and spin indices,
respectively. 
The key structural feature of FeSC is
an iron atom enclosed in a tetrahedral cage of 
pnictogen or chalcogen
atoms. The tetrahedra are packed
in a checker-board arrangement, with a 
unit cell containing two iron atoms, separated by a common center of
inversion, satisfying this requirement. We now show how tRVB predicts
a condensate with the above properties.

In the parent compound of the FeSC, each tetrahedron
contains two electrons within the three 
$xz,yz$ or $xy$ orbitals of the $t_{2g}$ level, Hund's coupled
into a  $S=1$, $L=1$ manifold. Consider the ``atomic'' limit of
an isolated iron tetrahedron. 
Each  pair of $t_{2g}$ orbitals 
shares a common direction, for instance, the $xz$ and $yz$
orbitals share a common $z$ axis, which in the presence of spin-orbit
coupling causes~\cite{Supplement} 
the  Hund's interactions 
to develop an orbitally selective
easy-plane anisotropy (Eq.~\ref{eq:AnisotropicIA}),
\begin{eqnarray}\label{eq:Hund1}
H_{I} &=& - 2\Big[ (J_{H}+J_{\rm A})\vec{ S}_{xz}\cdot \vec{S}_{yz}- 
2J_{\rm A} S^{z}_{xz}S^{z}_{yz})\nonumber\\
&&\hspace{2.5cm} + (\hbox{cyclic permutations)} \Big].
\end{eqnarray}
Each of the three interaction terms stabilizes a triplet pair with zero spin component along a quantization axis (``d-vector'') normal to 
its easy-plane (See Fig. \ref{fig:tRVBIllust}c), thus the xz and xy orbitals
have d-vector $\hat d=\hat  x$. 

With the convention $a \in \{xz,yz, xy \} =
\{1,2,3 \}$, 
the projected angular momentum operator within the $t_{2g}$ subspace 
is  $(L_a)_{bc}\equiv{-}i\epsilon_{abc}$. 
Defining the triplet pair creation operators
$\Psi^\dagger_{ab} \equiv \psi\dg ({L}_a \sigma_b)\bar \psi\dg $, $a,b
= {1,2,3}$, 
where 
$\bar \psi^\dagger \equiv i \sigma_2
(\psi^\dagger)^T$, 
Eq.~\eqref{eq:Hund1} can be written using summation convention as 
$H_{\rm I} =  - g_{ab}
\Psi\dg_{ab}\Psi_{ab}, 
$
with $g_{ab}=\frac{1}{4}(J_{H}+ J_{\rm A}\delta_{ab})$.
In this way, we see that
an anisotropy $J_{\rm A}>0$ splits off a ground-state manifold 
of triplet pairs in which the orbital angular momenta and the spin
quantization axis are aligned, $\Psi^\dagger_{aa}\ket{0} = \psi \dg
(\sigma_{a}L_{a})\bar \psi \dg \ket{0}$. 

The spin-orbit coupling $H_{SL}=-\lambda\vec{ L}\cdot \vec{S}$ causes
the triplet valence bonds to resonate between orbitals, giving rise to a tRVB
ground state  $\ket{\text{tRVB}} = \sum_{ab} \Lambda^{ab}\Psi^{\dagger}_{ab}\vert
0\rangle $ (see
Fig.~\ref{fig:tRVBIllust}b). Note that within the t$_{2g}$ multiplet,
the projected spin orbit interaction has a reversed coupling constant,
with $\lambda>0$, favoring $L+S=2$ configurations. 
The structure of the resulting energy levels 
is modelled
by a crystal field Hamiltonian 
given by $H = -\lambda (\vec{L}\cdot
\vec{S})- \alpha (J_{x}^{4}+J_{y}^{4}+J_{z}^{4})+\eta J_{z}^{2}
$, where $J=S+L$ is the total angular momentum, $\alpha \sim
J_{A}$, while $\eta $ quantifies the 
tetragonal anisotropy of the environment. 
The 
simplest tRVB 
ground-state, where  ${\Lambda}_{ab}=\delta_{ab}
$ is a unit matrix,
develops for the wrong sign of the spin-orbit coupling
$\lambda<0$. Two other tRVB states with
$\Lambda_{ab}= {\rm diag}
(1,-1,0)$ and 
$\Lambda_{ab}={\rm diag}(1,1,-2)$
are stabilized for $\lambda>0$,
\cite{Supplement}, where the latter becomes the unique ground-state
in the presence of a tetragonal anisotropy $\eta >0$, see Fig.~(\ref{fig:tRVBIllust}b).

\begin{figure}
\includegraphics[width=\linewidth]{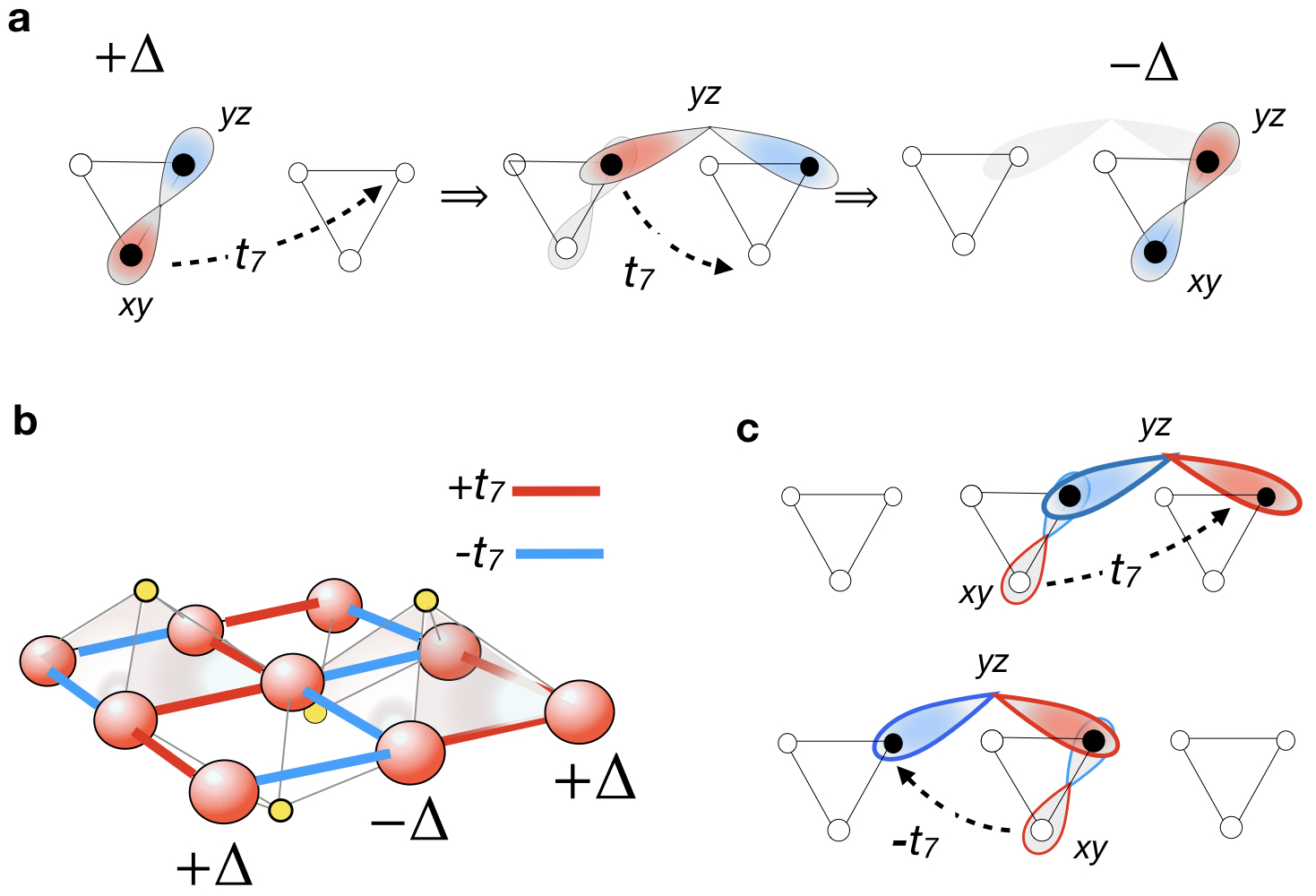}
\caption{\small Schematic showing a) how 
tunneling of a triplet valence bond between two iron atoms leads to
``tumbling'' motion that reverses the onsite triplet pair amplitude
$\Delta$ on
neighboring iron atoms, b) the 
alternation in the sign of 
inter-orbital hopping $t_7$ and onsite triplet pairing,
c) how the asymmetric left and right
tunneling 
permits triplet pairs to align in the same
direction between sites, allowing them to coherently condense into a
p-wave state on the
Fermi surface.\label{figy3}}
 \end{figure}

When the tetrahedra are brought together to form a conductor, 
charge fluctuations allow the escape of atomic triplet pairs into the
conduction sea.  
We shall assume that the interactions present in the
isolated tetrahedra are preserved in the metallic state that now
develops. 
Imagine a lattice where the $xy$ orbitals
are weakly hybridized with the $xz$/$yz$ orbitals at
neighboring sites (we denote this amplitude as $t_7$). An onsite valence bond
between an $xz$ and $xy$ orbital can tunnel to the neighboring site in a
two step process: 
an $xz$ electron 
first hops to a neighboring $xy$
orbital, forming an intersite, intraorbital triplet pair, after which the $xy$ electron 
follows suit and hops onto the neighboring site to reassemble
the intra-atomic triplet bond.  In fact, the electrons 
can tunnel in
either order and the resulting tumbling motion of
the tRVB causes its amplitude to alternate at neighboring sites. If
this process becomes coherent, it leads to a staggered anomalous triplet 
pairing amplitude (see eq \ref{invert})
$\Delta (\bx )= - \Delta (-\bx )
$
as envisioned in \cite{AndersonTriplet} (see Fig. \ref{figy3}a).
For this motion to be sustained coherently, there must be
two atoms per unit cell. To understand how this works in the FeSC, we
note there is
an additional non-symmorphic symmetry \cite{LeeWen2008}, under which the lattice is invariant
under a glide and mirror reflection through the plane. 
The opposite parities of the $xy$ and 
$xz$/$yz$  orbitals under glide reflection, means that the inter-orbital
tunneling amplitude $t_{7}$ \emph{alternates} (see Fig. \ref{figy3}b). 
When the $xz$/$xy$ and $yz$/$xy$ pairs
tunnel left, or right into the conduction sea, they do so with
opposite amplitudes, causing the intersite, intraorbital triplet pairs to 
coherently condense in the same direction. This permits the phase-alternating tRVB
pairs to coherently escape onto the Fermi surface (see Fig. \ref{figy3}c), activating a logarithmic Cooper divergence
in the pair susceptibility. 
The non-symmorphic symmetry of the FeSC allows us
to absorb the  staggered \red{hopping} into
a staggered gauge transformation
of the $xz$/$yz$ orbitals~\cite{DaghoferDagotto2010}, $\psi_{xz/yz}
({\bf j})\rightarrow (-1)^{j_x+j_y}\psi_{xz/yz} ({\bf j})$. This transformation unfolds the Brillouin zone and allows 
to treat each
iron atom on an equal footing.

Following \cite{Anderson1987} we introduce the \red{simplest} tRVB wave function as the Gutzwiller projection of a BCS-like wave function
\begin{equation}
\vert\hbox{tRVB}\rangle = \hat P_{G} \prod_{\mathbf k} \text{exp}\left (\psi^\dagger_{\mathbf k} [\vec{\mathcal L}(\mathbf k)\cdot\vec\sigma] \bar{\psi}^\dagger_{-\mathbf k} \right) \ket{0}. \label{eq:tRVBCrystal}
\end{equation}  
 Here $P_G$ is the Gutzwiller projector to $n<2$ electron per site.
The functions $\vec{\cal L}=\sum_g\vec\Lambda_g(\bk)\lambda_g$ with $g=1...8$
can be expanded in the eight-fold space of Gell-Mann matrices which span the
t$_{2g}$ multiplet. The triplet character of the condensate means that
${\cal L} (-\bk )= -{\cal L}^{T} (\bk )$, so
the three anti-symmetric $\lambda_g \in \{L_a\}_{a = 1}^3$ matrices 
combine with even parity functions $\Lambda_{s}
(\bk)=\Lambda_{s} (-{ \bk})$ to describe the onsite,
orbitally antisymmetric pairing, while the five
symmetric $\lambda_g$, 
combine with odd-parity p-wave functions $\Lambda_{a}
(\bk)=-\Lambda_{a} (-\bk)$, to describe the tRVBs that have
escaped to the Fermi surface. 

To calculate the properties of the tRVB wavefunction,  we adopt a Gutzwiller
mean field approach, assuming that the action of the microscopic
Hamiltonian beneath the projection operator $P_{G}$ can be modelled by
\red{an appropriate renormalization of hopping matrix elements in a}
mean-field Hamiltonian. \red{A microscopic rationale for these
renormalizations can be obtained from a slave boson treatment of the
unprojected Hamiltonian, along the lines of RVB
theory~\cite{KotliarLiu1988,KomijaniProgress}. Here we concentrate on the
weak-coupling Cooper instability that arises from the renormalized
Hamiltonian. }
Motivated by our discussion of the isolated tetrahedron, we now rewrite
the Hund's interaction,  Eq.\,\pref{eq:Hund1} in the form of a BCS theory
\be
H_I=\sum_{\mathbf x,ab}\Big[\frac{1}{g_{ab}}\bar{\Delta}_{ab}{{\Delta}_{ab}}+( \Psi^\dagger_{ab}{\Delta}_{ab}+h.c.)\Big ].\label{eq6}
\ee
For $t_{2g}$ materials, the states at the Fermi surface are composed
of three component Bloch wave functions $\vec u_{n, \mathbf k}$ which
are eigenstates of the kinetic term $H_{\rm kin}(\mathbf k) \vec u_{n,
\mathbf k} = \epsilon_{n}(\mathbf k) \vec u_{n, \mathbf k}$. On the
Fermi surface, the band-diagonal matrix element of the 
gap function is given by $\vec{d}_{n\bk }\cdot \vec{\sigma }$, where
the d-vector is $d^{a}_{n\bk }\equiv 
\Delta_{ab} (\vec u_{n, -\mathbf k}^T  L_b 
\vec u_{n, \mathbf k}) =-i \Delta_{ab} (\vec u_{n, -\mathbf k} \times \vec u_{n,
\mathbf k})_{b}$. The d-vector vanishes if the Bloch wave function 
$\vec u_{n-\bk }= \vec u_{n\bk }$ is symmetric, since
$\vec u_{n, \mathbf k} \times \vec u_{n,\mathbf k}=0$. 
Fortunately, the non-symmorphic 
character of the lattice mixes the $xy$ and $xz$/$yz$
orbitals, so that $\vec u_{n\bk }\neq \vec u_{n-\bk}$, 
which allows the d-vector to be finite. 

The simplest mean-field theory, corresponding to $\Delta_{ab} =\Delta {\rm diag} (1,1,-2) $,
models the iron-based superconductors as a two dimensional conductor
with Hamiltonian
\begin{eqnarray}\label{eq:MF}
H_{\rm BCS}&=&{\frac{|\Delta |^{2}}{g}+\frac{1}{V}\sum_{\bk }\tilde\psi_{\bk }\dg 
\Big[H_{\rm kin}(\mathbf k)\tau_{3}}\nonumber \\
&&\hspace{1.5cm}{+ \Delta (\sigma_1L_1+\sigma_2L_2-2\sigma_3L_3)\tau_{1} \Big]
\tilde\psi_{\bk }}.\qquad
\end{eqnarray}
Here $\tilde\psi_{\bk}$ is a Nambu spinor in the space of orbital, spin and
charge (isospin) space.  The pairing term 
$(\sigma_1 L_1+\sigma_2 L_2) \tau_{1}$ term 
retains the essential tRVB pairing components that mix the $xy$ and
$xz$/$yz$  orbitals {at the Fermi surface and is sufficient to gap out
the Fermi surface. {In our two dimensional model
the component $\sigma_3L_3\tau_1$ has no weak-coupling support on the
Fermi surface but induces inter-band pairing between $xz$ and $yz$
orbitals~\cite{VafekChubukov2017}. }
The term}
\begin{equation}\label{eqHam}
H_{\rm kin}({{\bf k}})=
\epsilon_{\bk } +\vec{\epsilon}_{\bk}\cdot\vec\gamma =
\matc{cc|c}{a_{\bk } & g_{\bk } & i p_{k_{x}}\cr
g_{\bk } & b_{\bk } & ip_{k_{y}} \cr
\hline
-ip_{k_{y}}& -i p_{k_{x}}& e_{\bk }},
\end{equation}
describes the band-dispersion \cite{DaghoferDagotto2010}, 
where $a_{\bk } = 2t_1c_x+2t_2c_y+4 t_{3}c_{x}c_{y}-\mu$, 
$b_{\bk }= 2t_2c_x+2t_1c_y+4t_3c_xc_y-\mu$,
$g_{\bk }= 4 t_{4} s_{x}s_{y}$, $p_{k_{x}}= 2t_{7}s_{x}+4
t_{8}s_{x}c_{y}$, 
$p_{k_{y} }= 2t_{7}s_{y}+4 t_{8}s_{y}c_{x}$
and $e_{\bk }= 2t_{5} (c_{x}+c_{y})+ 4 t_{6}c_{x}c_{y}-\mu + \delta_{xy}$, and we have employed the short-hand notation
$c_{l}\equiv \cos k_{l}$ and $s_{l}= \sin k_{l}$ (l=x,y).

Although the pairing in this mean-field theory is uniform,
if we undo the gauge transformation of the $xz$/$yz$  states, the onsite
pairing between the $xy$ and $xz$/$xy$ states acquires the staggered
behavior predicted by Anderson. 
Remarkably, even though this order parameter is staggered, it
induces a gap on the Fermi surface, with a 
pair susceptibility that is logarithmically divergent at low temperatures.

\begin{figure}
\includegraphics[width=\linewidth]{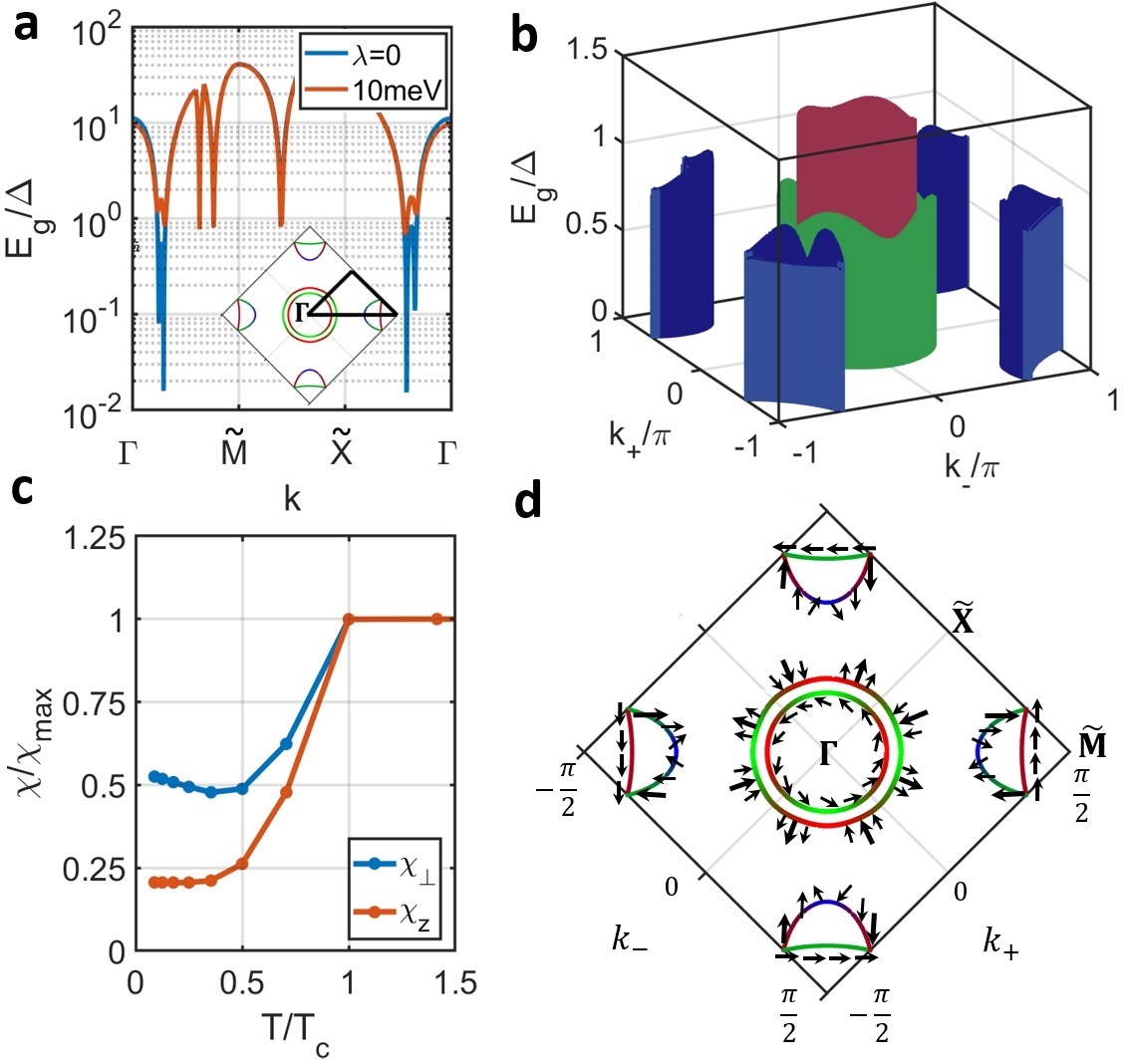}
\caption{\small a) The size of the gap along a cut passing high-symmetry points in the Fermi surface (FS), for $\Delta=6.2$meV \red{for $\lambda_{SO}=0$ and $\lambda=10$meV}. The inset shows the folded Brillouin zone with with $k_\pm=(k_x\pm k_y)/2$ and $\tilde X=(\pi,0)$ and $\tilde M=(\pi/2,\pi/2)$. 
\red{b) The size of the gap on the FS for $\Delta=6.2$meV and $\lambda_{SO}=10$meV}. \red{c}) The normalized spin-susceptibility at the transition for $\Delta=6.2$meV and $\lambda_{SO}=$ \red{10}meV~\cite{BorisenkoZhigadlo2016}. \red{d}) The winding of the $\vec d({\bf k})$ vector along the FS for $\lambda_{SO}=0$ \red{illustrates p-wave ($E_u$) pairing}. Note that $\vec d$ vector is entirely in the plane in this case.
}
\label{fig:fig3}
\end{figure}

\red{Fig{s}. \ref{fig:fig3}~a,b} display the spectrum calculated from the mean-field
theory Eq.~\eqref{eq:MF} using tight binding parameters of
Ref.~\cite{DaghoferDagotto2010} and $t_8=-t_7/3$. \red{The ground-state
develops} an anisotropic, yet full gap on the Fermi
surface \red{which becomes increasingly isotropic with the introduction 
of spin-orbit coupling}.  
%
Historically, 
the observation of a full gap 
\cite{SprauDavis2017,KushnirenkoBorisenko2018,HashimotoShin2018}
and
the presence of a finite  Knight shift in all field directions
led to an \red{early} rejection 
of the idea of triplet pairing in FeSC. 
\red{ However}, the calculated Knight-shift, obtained by summing both Fermi surface and inter-band
components of the total spin and orbital susceptibility (Fig. \ref{fig:fig3}\red{c}), 
\red{shows a} marked loss of spin susceptibility for
all field directions, \red{in accord with experiment.} 
We note that in a two dimensional model, 
the staggered hopping $t_{7}$ that delocalizes the pairs
is only present in the basal plane. 
When motion in the c-axis is included, the additional staggered
hopping along the c-axis will now hybridize the $xz/yz$ orbitals,
introducing an additional $p_{z}$ component to the condensate,
\red{further reducing the predicted anisotropy.}

Various other aspects of the tRVB theory of pairing in FeSC 
\red{deserve discussion.
First, since the Hund's triplet  pairing occurs 
locally on the iron atom, (unlike, 
$s_\pm$
pairing)}, tRVB accounts for intra-atomic Coulomb
repulsion without relying on a cancellation between electron and hole
pockets
~\cite{KoenigColeman2019}. \red{
Second, because this pairing is local, it is expected to be moderately robust against
the pair breaking effects of impurity scattering.  Microscopically,
disorder generates non-zero vertex corrections to the local pair
which partially cancel the disorder induced self
energy~\cite{Supplement}, thereby reducing the pair-breaking
effets of disorder. 
Third}, there are multiple sign
changes of the triplet d-vectors on and in between the various
Fermi surfaces (Fig.\,\ref{fig:fig3}\red{d}). The finite winding number of the d-vector
around each pocket may lead to interesting topological behavior. At
the same time the relative sign 
between d-vectors on electron and hole
pockets 
\red{gives rise to quasiparticle coherence factors  which closely resemble those
of an $s_{+-}$ superconductor with important consequences for quasiparticle
interference (QPI)~\cite{HanaguriTakagi2010,ChiPennec2014} and neutron spin resonance measurements .  Specifically, the dominant Fermi surface contribution to the antisymmetrized tunneling density of states at wave vector ${\bf q}$ is proportional to the Fermi surface (FS) average $\langle 1- \hat d_{n}({\bf k}+{\bf q}) \cdot \hat d_{m}({\bf k}) \rangle_{{\bf k} \in \rm FS}$, with $\hat d= \vec d/\vert \vec{d} \vert$. Features in this observable were previously interpreted as evidence for $s_\pm$ pairing, but our estimate suggests that tRVB is also consistent. A more detailed expression and a discussion of a similar feature on the} subgap
spin-resonance~\cite{ChristiansonGuidi2008} \red{are relegated to Ref.~\cite{Supplement}}.

\red{A key feature of tRVB is the prediction 
that Hund's pairing will give rise to a staggered
superconductor. The manifestation of this state in FeSC and
other candidate materials, 
would be most naturally detected as a spatial modulation
in the relative phase of the Josephson current 
measured in a scanning tunneling Josephson microscope,
using two superconducting STM tips of the same tRVB
material. 
The alternating superconducting phase is predicted to} lead to a
staggered $\pi$-junction behavior as the
tip is swept across the material~\cite{Supplement}.

Finally, we mention the possible relevance of tRVB
to other superconductors of current interest. 
The recent discovery of the 
heavy-fermion UTe$_2$, which has an even number of uranium atoms per
unit cell, with likely 
triplet superconductivity \cite{Ran2019} is one promising
example. Another intriguing candidate material is
magic angle double bilayer graphene, where the valley degrees of freedom play the
role of orbitals, giving rise to Hund's coupled 
interorbital triplet pairing~\cite{ScheurerSachdev2019}
on a  moir\'e superlattice.\\

\textit{Acknowledgments:} The authors gratefully acknowledge discussions with
Po-Yao Chang.
Piers Coleman and Elio K\"onig are supported by DOE Basic Energy
Sciences grant DE-FG02-99ER45790.  Yashar Komijani was supported
by a Rutgers Center for Materials Theory postdoctoral fellowship. All authors contributed equally to this work.

\begin{widetext}
\section*{Supplementary Information}

These supplemental materials include a section on the preservation of
triplet (tRVB) states under time evolution (Sec.~\ref{App:OverComplete}), 
a discussion of the role of symmetries and impurity scattering (Sec.~\ref{App:FeSC:NoGoTheorem}), and a study of observables for the iron-based superconductors, including the local density of states, Knight shift and a proposal for the detection of the staggered superconducting phase using scanning tunneling Josephson microscopy  (Sec.~\ref{App:FeSC:Observables}).\\

\section{Preservation of $t$RVB states under time evolution}
\label{App:OverComplete}
\begin{figure}[b]
\includegraphics[width =0.5\textwidth]{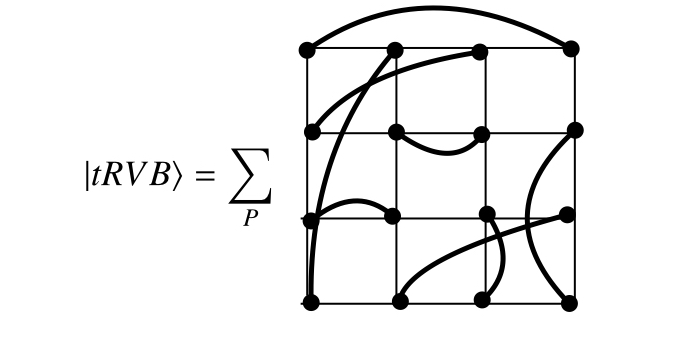}
\caption{Bond configurations in a tRVB
wavefunction.
}
\label{fig1}
\end{figure}

The concept of the tRVB state relies on the observation
that the ground-state
xy-anisotropic Ferromagnet, with Hamiltonian $H=\sum_{(i,j)}H_{ij}$,
where  
\begin{equation}\label{}
H_{ij}= - J (\vec{S}_{i}\cdot \vec{S}_{j})+ \Delta  J
S^{z}_{i} S^{z}_{j}
\end{equation}
is a resonating valence
bond state of triplet pairs (see Fig.~\ref{fig1}), given by  a
weighted sum over bond configurations 
\begin{eqnarray}\label{l}
|{\rm tRVB}\rangle	  &=& \sum_{P}A_{P} \vert P\rangle, \cr
\vert P \rangle &=& \prod_{(i,j)\in P}
\vert (i,j)\rangle .
\end{eqnarray}
Here $A_{P}$ is the amplitude for a given configuration $\vert
P\rangle $ of triplet valence bonds (tVBs) and 
$\vert (i,j)\rangle \equiv
(\vert \uparrow\rangle _{i}\vert \downarrow\rangle _{j}+\vert
\downarrow\rangle _{i}\vert \uparrow\rangle _{j})/\sqrt{2}$ is an
$m=0$ triplet valence bond formed between sites $i$ and $j$. 
In contrast to its singlet cousin, which has been extensively studied,
the properties of tRVB ground-states are largely unexplored.
One of the important points that was learned from the study of
RVB ground-states, is that even nearest neighbor, ``dimer'' coverings 
can exhibit off-diagonal long range antiferromagnetic order (see
eg. \cite{Albuquerque12}). Similar behavior is expected for the dimer
tRVB state. 


The consistency of tRVB theory requires that  the  action of the
Hamiltonian on any configuration of the triplet valence bonds (tVBs)
is closed within the space of tVBs, i.e
that the action of the Hamiltonian on a given bond, $H_{ij}$ lies exclusively
within the space of states $\left\{ \vert P\rangle  \right\}$, so that 
$H_{ij}\vert P\rangle = \sum_{P'} \vert P'\rangle h^{P'P} (ij)$. 
We can rewrite the isotropic part of the Hamiltonian
in terms of the spin exchange  operator $P_{ij}$, 
\begin{equation}\label{}
H_{ij} = - ( {J}/{2})P_{ij}+\Delta JS^{z}_{i} S^{z}_{j}+J/4. 
\end{equation}
The action of $P_{ij}$
permutes the ends of the valence bonds, so it is closed 
within the Hilbert space of tVBs, 
however  the action of the 
additional Ising component $H^{I}_{ij}= \Delta J S^{z}_{i}S^{z}_{j}$ 
needs to be considered
with care. 

There are two configurations of the tVBs to consider (Fig. \ref{fig2}). 
If there is a triplet valence
bond between $i$ and $j$, then  it is unaffected by the Ising term
$S^{z}_{i}S^{z}_{j}
\vert (i,j)\rangle  
= -\frac{1}{4}
\vert (i,j)\rangle 
$(Fig. \ref{fig2}a).
If however, there is no bond between sites $i$ and $j$, then we 
must have  two separate tVBs, one linked to site $i$, the other to
site $j$. In this situation, 
the Ising term 
has
the effect of converting two triplet bonds ending at $i,j$ into two 
{\sl singlet bonds} which at first sight, suggests
 that the space of tVBs is not closed 
under the action of $H_{ij}$. However, we now demonstrate that the space is closed. To this end we carefully take into account the overcompleteness of the
basis of valence bonds states. (Fig. \ref{fig2}b).
\begin{figure}[h]
\includegraphics[width =\textwidth]{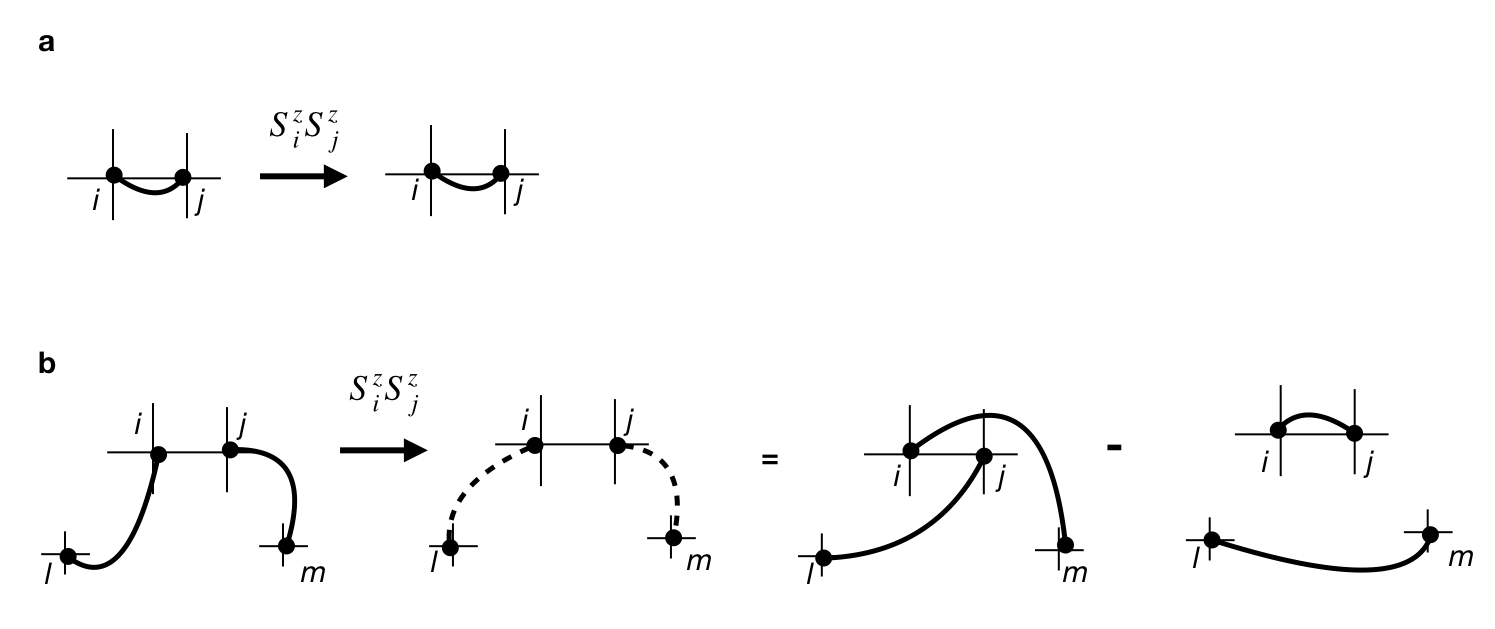}
\caption{
a) The action of the Ising part of  $H_{ij}$
on a single bond $(i,j)$ leaves it invariant. b) The action of the
Ising part of $H_{ij}$ on two bonds linked to $i$ and to $j$
converts them to singlet bonds (dashed lines), which can then be
re-written as a sum of two tVB configurations.
}
\label{fig2}
\end{figure}

If $(i,l)$ and
$(j,m)$ are two triplet valence bonds from other sites $l$ and $m$
which terminate at $i$ and $j$ respectively, then 
\begin{equation}\label{}
S^{z}_{i} S^{z}_{j} \ \vert (i,l)\rangle  \vert (j,m)\rangle 
= \frac{1}{4}\vert [i,l]\rangle  \vert [j,m]\rangle ,
\end{equation}
where we have employed the commutator notation  $\vert [i,j]\rangle \equiv 
(\vert \uparrow\rangle _{i}\vert \downarrow\rangle _{j}- \vert \downarrow\rangle _{i}\vert \uparrow\rangle _{j})/\sqrt{2}$
to describe singlet
RVBs. At first sight, this implies that the Ising terms will lead to a
mixture
of singlet and triplet bonds. 
However,  the overcompleteness of the RVB
representation, allows us to represent these 
two singlet bonds as a superposition of triplet bonds (see Fig. \ref{fig2}b).  
Direct algebraic expansion confirms that 
\begin{equation}
\vert [1,2][3,4]\rangle  = \ket{(1,4)}\ket{(2,3)}-\ket{(1,3)}\ket{(2,4)}.
\end{equation}
which guarantees that the tRVB 
manifold of states that is closed under the time-evolution.

\section{Symmetry constraints on $\rm\bf t$RVB and impurity scattering}
%
\label{App:FeSC:NoGoTheorem}

One of the key aspects of the tRVB theory, is the ability of local
triplet pairs, formed within an atom, to escape and form a coherent
condensate on the Fermi surface.  Here we illustrate how the
constraints of inversion symmetry in the FeSC allow this process to
take place.

From Eq. (8) of the main text, the tRVB BCS Hamiltonian is 
\begin{equation}
H = H_{\rm kin}+H_{P}=\sum_{\bf k} \tilde \psi_{\bf k}^\dagger \left [H_{\rm kin}({\bf
k}) \tau_3 +  \Delta_{ab} L_{a} \sigma_{b} \tau_{1} 
\right ] \tilde \psi_{\bf k}.
\end{equation}
where the sum is over half the Brillouin zone, to avoid
double-counting.
Following the notation of the main text, we denote the multi-orbital
Bloch wavefunctions by $\vec u_{n,{\bf k}}$, which are the eigenstates of
the  tight-binding Hamiltonian,  $H_{\rm kin} (\bk ) \vec
u_{n,{\bf k}} = \epsilon_{n}({\bf k}) \vec u_{n,{\bf k}}$. 
As we show in (\ref{deriv})
the band diagonal pairing
matrix elements of the superconducting gap are then related
to the eigenvectors $\vec u_{n,{\bf k}}$ according to 
\begin{equation}\label{}
\langle 0 \vert H_{P}
\vert n,-\bk, \alpha ;n\bk \beta \rangle 
= 
\vec{d}_{n} (\bk )\cdot (-i\sigma_{2}\vec{\sigma })_{\alpha \beta }.
\end{equation}
Here $\vert n,-\bk, \alpha ;n\bk \beta \rangle 
= 
a\dg_{n\bk  \beta}a\dg_{n,-\bk \alpha }\vert  0 \rangle $ is a
triplet pair of electrons in the n-th band
and the d-vector
\begin{equation}\label{eq:GapMatrixElement}
\vec{d}_{n} (\bk ) = - \vec{d}_{n} (-\bk) = i (\vec{u}_{n,\bk }\times \vec{u}_{n,-\bk
})
\cdot\underline{\Delta},
\end{equation}
where $[\underline{\Delta }]_{ab}=\Delta_{ab}$ is the onsite gap
function.

A finite magnitude  of the vector
$\vec{u}_{n,\bk }\times \vec{u}_{n,-\bk}$ plays a crucial role, for it
allows the onsite pairing to migrate to the Fermi
surface, giving rise to a gap $\Delta_{n} (\bk )\sim |\vec{d}_{n}(\bk
)|$ that grows linearly with the order parameter $\underline{\Delta}$. 
Moreover,  the linear growth of the Fermi surface gap with the order
parameter guarantees that the pair susceptibility 
will acquire a logarithmic divergence in temperature, driving a
Cooper instability at arbitrarily weak coupling. 
Conversely, if $\vec{u}_{n,\bk }\times \vec{u}_{n,-\bk}=0$ 
is zero, the pairing is entirely inter-band in character,
there is no weak-coupling instability and the superconducting gap 
does not grow linearly with the order parameter. 

In section (\ref{conditions}) we discuss the conditions under which 
$\vec{u}_{n,\bk }\times \vec{u}_{n,-\bk}$  is finite. 

\subsection{Derivation of the matrix element}\label{deriv}

The pairing component of the Hamiltonian can be written out as
\begin{equation}\label{}
H_{P}  = \sum_{\bk } \left[\Delta_{ab}\bar \psi_{-\bk }^{T}
(L_{a}\sigma_{b})\psi_{\bk }+ {\rm H. c} \right],
\end{equation}
where band and spin indices 
denoted $\bar
\psi _{-\bk }^{T}= \psi^{T}_{-\bk } (-i\sigma_{2})$.  To transform this into the band-basis, we 
note
that the components of $\vec u_{n,\bk }$ can be written in Dirac notation as
the overlap between the orbital and band bases $(\vec u_{n, \bk })^{\alpha }
=\langle \bk \alpha |n\bk \rangle$, where $\alpha $ is the orbital
index, and $n$ the band index. Now using completeness, 
the relationship between ``bras'' in the two bases is  $\langle  \bk  \alpha \vert
= \sum_{n}\langle \bk  \alpha \vert n \bk \rangle
 \langle n \bk  | $, and
since destruction operators transform like ``bra''s, 
it follows that $\psi_{\bk  a} = \sum_{n}\langle \bk  \alpha \vert n \bk \rangle
a_{n\bk }$,  or in terms of $\vec u_{n, \bk }$ and band annihilation operator $a_{n\bf k}$, 
\begin{equation}\label{}
\psi_{\bk }= \sum_{n}\vec u_{n, \bk }a_{n\bk }, \qquad 
\bar \psi^{T}_{-\bk } = \sum_{n}
\vec u^{T}_{n, \bk }\bar a^{T}_{n,-\bk },
\end{equation}
where $\bar a^{T}_{n,-\bk }= a^{T}_{n,-\bk } (-i\sigma_{2})$.
Using these relationships, we can re-write the pairing term in the
band-basis as
\begin{eqnarray}\label{HP}
H_{P}&=&\sum_{\bk,a,b,m,n }\left[\Delta_{ab} (\vec u^{T}_{m\bk }L_{a}\vec u_{n, \bk
}) \bar a^{T}_{m,-\bk } \sigma_{b}a_{n\bk}+ {\rm H.c}
\right]
\cr
&=&  \sum_{\bk ,m,n}\left[ 
 \bar a^{T}_{m,-\bk } (\vec{ d}_{mn} (\bk )\cdot\vec\sigma)a_{n, \bk}+ {\rm H.c}\right],
\end{eqnarray}
where  the d-vector
\begin{equation}\label{}
[d_{mn} (\bk )]_{b}= \sum_{a}
 (\vec u^{T}_{m\bk }L_{a}\vec u_{n, \bk
})\Delta_{ab}.
\end{equation}
Now since $[L_{a}]_{bc}= - i \epsilon_{abc}$, it follows that 
\begin{equation}\label{}
(\vec{d}_{mn} (\bk ))_{b} =
 -i (\vec{u}_{m,-\bk }\times\vec{u}_{n, \bk })_{a}\Delta_{ab}\equiv 
\left[ -i (\vec{u}_{m,-\bk }\times\vec{u}_{n, \bk })\cdot \underline{\Delta}
\right]_{b},
\end{equation}
where we have used a matrix notation to write $\underline{\Delta
}_{ab}\equiv \Delta $.
The ability of local pairs to migrate onto the Fermi surface depends
on the band-diagonal component of this matrix element, 
\begin{equation}\label{}
\vec{d}_{n} (\bk ) = i (\vec{u}_{n, \bk }\times \vec{u}_{n,-\bk
})
\cdot\underline{\Delta},
\end{equation}
where we have denoted $\vec{d}_{n} (\bk )\equiv \vec{d}_{nn} (\bk )$.
Notice that because the cross-product is antisymmetric, the diagonal
d-vector is odd parity in momentum,
 $\vec{d}_{n} (\bk )= i (\vec{u}_{n,-\bk }\times \vec{u}_{n,\bk
})
\cdot\underline{\Delta} = - i (\vec{u}_{n,\bk }\times \vec{u}_{n,-\bk
})
\cdot\underline{\Delta}= -\vec{d} (\bk )$.

Let us  now compute the 
the amplitude to 
destroy a triplet Cooper pair out of the vacuum,  $\langle 0 \vert H_{P}
\vert n,-\bk, \alpha ;n, \bk \beta \rangle $, where 
where $\vert n,-\bk, \alpha ;n, \bk \beta \rangle 
= 
a\dg_{n\bk  \beta}a\dg_{n,-\bk \alpha }\vert  0 \rangle $. 
Substituting this into Eq(\ref{HP}) and explicitly exposing the spin
indices, 
we obtain
\begin{eqnarray}\label{l}
\langle 0 \vert H_{P}
\vert n,-\bk, \alpha ;n\bk \beta \rangle 
&=& \langle 0\vert  H_{P}
a\dg _{n\bk \beta }
a\dg _{n-\bk \alpha}\vert 0\rangle \cr
&=&\sum_{l,m,\gamma,\delta } 
\langle 0 \vert a_{l,-\bk  \gamma}
\left( 
\vec{d}_{lm} (\bk ) \cdot(-i\sigma_{2}\vec{\sigma })_{\gamma\delta
}\right)
a_{m\bk \delta }
a\dg _{n\bk \beta }
a\dg _{n-\bk \alpha}\vert 0\rangle \cr
&=&
\vec{d}_{nn} (\bk ) \cdot(-i\sigma_{2}\vec{\sigma })_{\alpha\beta }\cr
&\equiv &
\vec{d}_{n} (\bk ) \cdot(-i\sigma_{2}\vec{\sigma })_{\alpha\beta }.
\end{eqnarray}

\subsection{Conditions for a finite gap}\label{conditions}
As a second step we discuss the symmetry enforced properties of $\vec
u_{n,\bf k}$ in our two dimensional model of FeSe, identifying the
non-trivial representation of the two-dimensional inversion symmetry,
resulting from the non-symmorphic crystal structure as an origin of
the finite Fermi surface support of the gap. The two-dimensional
inversion symmetry,  according to which, $H_{\rm kin}({\bf k}) =
M^{-1} H_{\rm kin}(-{\bf k}) M$ ($M$ is unitary), implies  that 
$\vec u_{n, -{\bf k}} = M \vec u_{n, \bf k},$.
Thus if the representation of inversion symmetry is trivial, so that 
$M = \mathbf 1$ 
the Fermi surface d-vector vanishes
\begin{equation}\label{eq:GapMatrixElement}
\vec{d}_{n} (\bk ) = i (\vec{u}_{n, \bk }\times \vec{u}_{n, \bk
})
\cdot\underline{\Delta}=0,
\end{equation}
so that the onsite triplet pairing does not escape to the Fermi
surface. 
%
This is the essence of the observations by Anderson~\cite{AndersonTriplet}, Hotta and Ueda~\cite{Hotta04}.

For the specific case of the layered iron-based superconductors,
treated in a 2D model of a single plane, 
the three $t_{2g}$  orbitals $\ket{xz}$, $\ket{yz}$ and $\ket{xy}$
transform differently under the 2D inversion operation,
$(x,y,z)\to(-x,-y,z)$, which results in a non-trivial representation
$M=(-1,-1,1)$ of the 2D inversion. As a result, even in presence of
time-reversal symmetry (which implies $\vec u_{n,- \bk} = \vec u_{n,
\bk}^*$) the components of $\vec u_{n,\bf k}$ cannot all be real, resulting in a non-zero $\vec d$ vector,
\be
\vec u_{n,\bf k}=M\vec u^*_{n,\bf k} \Rightarrow \vec u_{n, {\bf k}}=\mat{iu_{n,xz}\\ iu_{n,yz} \\ u_{n,xy}}.
\ee
For the case $\Delta_{ab} = \Delta {\rm diag}(1,1,-2)$ considered in the paper,
\be
 \vec d_n({\bf k})=\Delta u_{n,xy}\mat{-u_{n,yz} \\ u_{n,xz} \\ 0}.
\ee
Note that the $\vec d$ vector is in the x-y plane and its value crucially depends on the $xy$ orbital admixture of the electrons at the Fermi surface. In other words, if the $xy$ orbital is localized there will be no triplet superconductivity. 

\subsection{Robustness against disorder}

A question which is related to the symmetries of the superconducting
gap regards the stability of $T_c$ against the inclusion of scalar impurities. 
In this section we outline a comparison of usual $s$-wave, usual $p$-wave, $s_{+-}$ and tRVB pairing and loosely follow the
textbook~\cite{AGD}. We consider a multiorbital
superconductor, which in the clean limit has Nambu-Gor'kov Green's function 
\begin{equation}
\mathcal G(i\nu, {\bf k}) = [i \nu - \mathcal  H({\bf k})]^{-1}.
\end{equation}

\begin{figure}[tp!]
\includegraphics[scale=1.25]{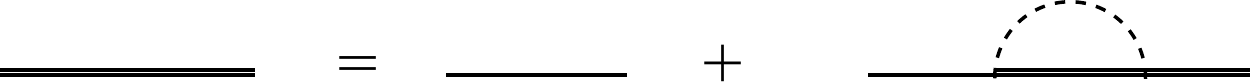}
\caption{Resummation of impurity scattering in non-crossing approximation using Nambu matrix Green's functions.}
\label{fig:Disorder}
\end{figure}

Here, $\mathcal H({\bf k}) = H_{\rm kin}(\bk) \tau_3 + \hat \Delta ({\bf k})\tau_1 $, where $\hat \Delta({\bf k})$ is a matrix in spin and orbital space, e.g. $\hat \Delta ({\bf k})= \sum_{a,b}\Delta_{ab} L_a \sigma_b $ in the tRVB case, $\hat \Delta ({\bf k})= \Delta \cos(k_1)\cos(k_2)$ for $s_\pm$ pairing and  $\hat \Delta ({\bf k})= \Delta \sum_{a=1,2}\sin(k_a) \sigma_a$ for ordinary $p$-wave superconductors. The diagrammatic, non-crossing resummation of impurity lines, Fig.~\ref{fig:Disorder}, of point like scatterers of strength $u_0$ and density $n_{\rm imp}$ leads to
\begin{eqnarray}
[\mathcal G^{-1}(i\nu, {\bf k}) - \Sigma (i\nu, {\bf k}) ] \boldsymbol{\mathcal G}(i\nu, {\bf k}) &=& \mathbf 1, \\
\Sigma(i\nu, {\bf k}) &=&  n_{\rm imp} u_0^2 \int_{\bf p} \tau_3\boldsymbol{\mathcal G}(i\nu, {\bf p}) \tau_3.
\end{eqnarray}
We use the notation $\int_{\bf p}=\int \frac{d^dp}{(2\pi)^d}$.
To establish the  stability of the superconductivity against disorder,
we need to investigate the persistance of the Cooper instability. Since this is an intraband phenomenon we consider the coupled equations
\begin{eqnarray}
[\mathcal G_n^{-1}(i\nu, {\bf k}) - \Sigma_n(i\nu, {\bf k}) ] \boldsymbol{\mathcal G}_n(i\nu, {\bf k}) &=& \mathbf 1, \\
\Sigma_n(i\nu, {\bf k}) &=& n_{\rm imp} u_0^2\sum_m \int_{\bf p} \vert \langle{u_{n,{\bf k}} \vert u_{m,{\bf p}}}\rangle \vert^2 \tau_3 \boldsymbol{\mathcal G}_m(i\nu, {\bf p}) \tau_3, \label{eq:Selfenergy}
\end{eqnarray}
where $\mathcal G_n(i\nu, {\bf k}) = [i \nu - \epsilon_n({\bf k}) \tau_3 - \Delta_n({\bf k})\tau_1]^{-1}$ and $\Delta_n({\bf k})=\langle{u_{n,{\bf k}} \vert \hat \Delta ({\bf k}) \vert u_{n,{\bf k}}}\rangle$,~{i.e.~only the intra-band part of the pairing is kept}. For orbital independent superconducting order parameters (e.g. ordinary s-wave), these expressions are exact. We will use the following matrix form
\begin{eqnarray}
\Sigma_n(i\nu, {\bf k}) = \left (\begin{array}{cc}
\Sigma^N_n(i \nu,{\bf k}) & \Sigma^A_n(i \nu, {\bf k}) \\ 
{[\Sigma^A_n(i \nu,{\bf k})]}^\dagger & - \Sigma^N_n(-i \nu,{\bf k}) 
\end{array}  \right) .
\end{eqnarray}

Now, following Abrikosov and Gor'kov, we seek a self-consistent solution using 
the low-frequency ansatz $\Sigma_n^N(i \nu, {\bf k}) = \tilde{\Sigma}_n^N({\bf k}) - i \nu \Gamma_{\nu,n}({\bf k})$ and $\Sigma_n^A(i \nu, {\bf k}) = - \Delta_n({\bf k}) \tilde{ \Gamma}_{\nu,n}({\bf k}) $, which can be justified a posteriori. Both self-energies are only weakly momentum dependent in the cases of interest. As usual~\cite{AGD}, the frequency independent part $\tilde{\Sigma}_n^N({\bf k})$ is absorbed into a renormalization of dispersion, chemical potential and crystal field and omitted from further considerations. Then, $\boldsymbol{\mathcal G}_n(i\nu, {\bf k})= [i \bar \nu_n - \epsilon_n({\bf k}) \tau_3 - \bar \Delta_n({\bf k})\tau_1]^{-1}
$ with $i \bar \nu_n=  i \nu (1+\Gamma_{\nu,n} ({\bf k}))$ and $\bar \Delta_n =
\Delta_n (1 +\tilde \Gamma_{\nu,n}({\bf k}) )$, where $1+\Gamma_{\nu,n}({\bf k})= Z^{-1}_{\nu,n}({\bf k})$
corresponds to the wavefunction renormalization of the Green's
function, while $\tilde{\Gamma}_{\nu,n}({\bf k})$ is the impurity
correction to the pairing vertex. 

The BCS equation of impure superconductors with $\hat \Delta ({\bf k}) = f({\bf k}) \hat \Delta $, where $f({\bf k})$ is a normalized form factor, is 
\begin{equation}\label{eq:BCSIMpure}
\frac{\hat \Delta}{g} =- T\sum_{\nu} \int_{\bf k} f({\bf k}) \mathbf F(i \nu,{\bf k})\simeq T\sum_{\nu,n} \rho_n(E_F)  \Big\langle \ket{u_{n,{\bf k}}} \frac{ f({\bf k}) \bar \Delta_n({\bf k})}{\sqrt{\bar \nu^2 + \bar \Delta_n({\bf k})^2}} \bra{u_{n,{\bf k}}}\Big\rangle_{\rm FS},
\end{equation}
 where $\langle \dots \rangle_{\rm FS}$ denotes the angular Fermi surface average and ${\mathbf F}(i \nu, {\bf k})$ denotes the anomalous Green's function. If ${\Gamma}_{\nu,n}({\bf k}) = \tilde{\Gamma}_{\nu,n}({\bf k})$, as it occurs in the simplest s-wave case, self-energy and vertex correction cancel in numerator and denominator of Eq.~\eqref{eq:BCSIMpure} and $T_c$ is unchanged (``Anderson's theorem''). In contrast, for $s_{\pm}$ pairing, $\tilde \Gamma_{\nu,n}({\bf k}) \ll \Gamma_{\nu,n}({\bf k})$ due to partial cancellation of contributions from electron and hole pockets in the anomalous self energy. Even more drastically, for ordinary single band p-wave triplet pairing, where $f({\bf k}) = \sum_{a = 1,2} \sin(k_a) \sigma_a$ is a matrix in spin space, $\tilde \Gamma_{\nu,n}({\bf k}) =0$ due to the  symmetries of the order parameter. Therefore, p-wave pairing is very susceptible to the presence of scalar impurities.

We now demonstrate that $\tilde \Gamma_{\nu n}({\bf k}) > 0$ for tRVB in iron based superconductors, despite the fact that the tRVB state is effectively p-wave on the Fermi surface. The self-consistent condition for the anomalous self-energy, Eq.~\eqref{eq:Selfenergy}, can be written as

\begin{eqnarray}
{\Delta_n({\bf k})} \tilde \Gamma_{\nu,n}({\bf k})
&=& - n_{\rm imp}{u_0^2} \langle u_{n,{\bf k}} \vert \Big [ \int_{\bf p} {\bf F}(i \nu, {\bf p}) \Big ] \vert u_{n,{\bf k}} \rangle
\end{eqnarray}

Therefore, a non-zero $\tilde \Gamma_{\nu,n}({\bf k})$ requires an onsite pairing amplitude, which is indeed a crucial aspect of tRVB theory.
For illustration, we consider the simplest tRVB state $\hat \Delta = \Delta \sum_{a = 1,2} L_a \sigma_a$. For this choice $\int_{\bf p}{\mathbf F}(i \nu, {\bf p}) \propto - \hat \Delta$ (keeping only intraband pairing) such that the matrix structure of Eq.~\eqref{eq:BCSIMpure} is indeed fulfilled in the leading logarithm approximation. We therefore find a momentum independent vertex correction which obeys

\begin{equation}
\tilde \Gamma_{\nu,n} = n_{\rm imp} u_0^2\sum_m \int \frac{d^dp}{(2\pi)^d} \frac{\vert \vec d_m({\bf p}) \vert^2 (1 +\tilde \Gamma_{\nu,m})/(2\Delta^2)}{  \bar \nu_n^2 + \epsilon_m({\bf p})^2 +\vert \vec d_m({\bf p}) \vert^2 (1 +\tilde \Gamma_{\nu,m})^2} >0.
\end{equation}
The finiteness of this quantity reflects the local character of the
tRVB pairing, and it is this feature that guarantees partial
protection with respect to elastic scattering. 

\section{Observables in iron based superconductors}
\label{App:FeSC:Observables}

For the BdG Hamiltonian $\mathcal H({\bf k}) = H_{\rm kin}(\bk) \tau_3 + \Delta_{ab} L_a \sigma_b \tau_1 $ 
the Matsubara Green's function can be expressed as
\begin{equation}
\mathcal G(i \epsilon_n, {\bf k}) \equiv [i\epsilon_n - \mathcal H({\bf k})]^{-1}. 
\end{equation}

In this section we concentrate on intraband pairing, use the notation $\Delta_n({\bf k}) =  \vec d_{nn}({\bf k}) \cdot \vec \sigma + d^{(0)}_{nn}({\bf k})$ and the multi index $(n,h)$, e.g. in the energy $E_{n,h}({\bf k}) =  \sqrt{\xi_n^2 + (d_{nn}^{(0)}+ h d_{nn})^2}$ (the quantum number $h = \pm 1$ denotes the eigenvalues $h d_{nn}$ of $\vec d_{nn} \cdot \vec \sigma$ and $\xi_n(\bk)$ is the dispersion in band $n$). For pure singlet or pure triplet states, $E_{n,h}({\bf k}) = E_{n,-h}({\bf k})$ and we can suppress the index $h$.
%
All observables in this section are computed for the tight-binding model of Ref.~\cite{DaghoferDagotto2010} and a gap function $\Delta (L_1 \sigma_1 + L_2 \sigma_2 - 2 L_3 \sigma_3)$.
The sign structure of the $\vec d$ vector on the Fermi surfaces of this tight binding model is summarized in Fig.~\ref{fig:Dvectors}.


\begin{figure}
\includegraphics[scale=.5]{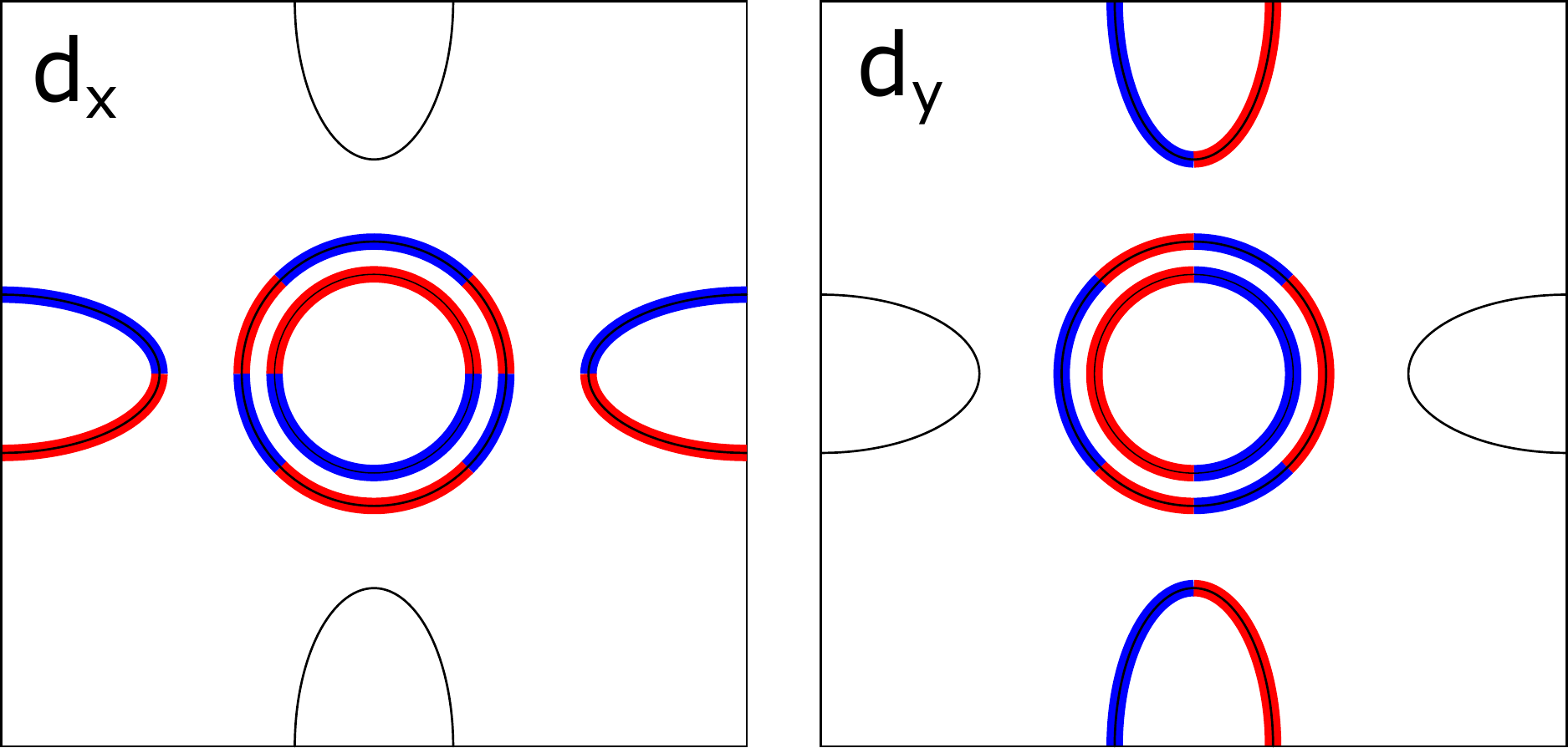}
\caption{Illustration of signs (red/blue color) of the components of d-vectors on the Fermi surfaces in the absence of SOC. Note that the $d_x$ ($d_y$) components approximately vanish on the $Y$ ($X$) electron pocket. The component $d_z = 0$ by symmetry.}
\label{fig:Dvectors}
\end{figure}

\subsection{Local density of states and QPI}

%

The even and odd frequency parts of 
the impurity contribution~\cite{MaltsevaColeman2009} to the local DOS in Fourier space $\delta \rho^{e/o}({\bf q}, z) = V({\bf q}) \Lambda^{e/o}({\bf q}, z)$ are 
\begin{eqnarray}
\Lambda^{e/o}({\bf q}, z) &=&  \frac{1}{2\pi} \text{Im} \int_{{\bf k}} \tr \left \lbrace \left (\begin{array}{c}
1 \\ 
\tau_3
\end{array} \right ) \mathcal G^A({\bf k}_+,z) \tau_3 \mathcal G^A({\bf k}, z)\right \rbrace \notag \\
&\simeq & \frac{2}{\pi} \sum_{n,m}\text{Im} \int_{\bf k} \frac{\vert \braket{{n,{\bf k}_+} \vert {m, {\bf k}}} \vert^2}{[z^2 - E_n({\bf k}_+)^2][z^2 - E_m({\bf k})^2]}\notag \\
&&\times \begin{cases} z (\xi_n({\bf k}_+) + \xi_m({\bf k})), & \text{ for }\Lambda^{e}({\bf q},z),\\ z^2 + \xi_n({\bf k}_+) \xi_m({\bf k}) - \vec d_{nn}({\bf k}_+) \cdot \vec d_{mm}({\bf k}),& \text{ for }\Lambda^{o}({\bf q},z),\end{cases}\label{eq:LocalDOSFinal}
\end{eqnarray}
where ${\bf k}_+ = {\bf k}+ {\bf q}$. Here, we considered predominant intraband pairing and the case of absent singlet pairing ($d_{nn}^{(0)} = 0$).
In the reverse case of absent triplet pairing ($\vec d_{nn} =0$), but present singlet pairing ($d_{nn}^{(0)} \neq 0$) we obtain the analogous result, i.e. Eq.~\eqref{eq:LocalDOSFinal} with the replacement $\vec d_{nn}({\bf k}_+) \cdot \vec d_{mm}({\bf k}) \rightarrow d_{nn}^{(0)}({\bf k}_+) d_{mm}^{(0)}({\bf k})$. 
At the Fermi surface ($\xi_n({\bf k}) = 0$) this integral is dominated by momenta at where $E_n({\bf k}_+) \equiv \vert \vec d_n({\bf k}_+) \vert$ and $E_m({\bf k}) \equiv \vert \vec d_m({\bf k}) \vert$ are equal and by frequencies $z$ which are on-shell. In order to illustrate this dominant physics in the main text, we replace $z \rightarrow \sqrt{\vert \vec d_n({\bf k}_+) \vert \vert \vec d_m({\bf k}) \vert}$, keeping in mind that the proper equation is Eq.~\eqref{eq:LocalDOSFinal}. 

When the STM bias voltage $eV$ is close to or below the superconducting gap, the sign of $\vec d_{nn}({\bf k}_+) \cdot \vec d_{mm}({\bf k})$ (or $d_{nn}^{(0)}({\bf k}_+) d_{mm}^{(0)}({\bf k})$ for the singlet case) in the numerator of Eq.~\eqref{eq:LocalDOSFinal} determines whether $\Lambda^{o}({\bf q}, eV)$ is enhanced (negative sign) or suppressed (positive sign) at a certain wavevector ${\bf q}$~\cite{HirschfeldMazin2015}. In particular, singlet $s_\pm$ pairing enhances $\Lambda^{o}({\bf q}, eV + i 0)$ at ${\bf q} \sim (\pi,0),(0,\pi)$ due to the relative sign $ d_{nn}^{(0)}({\bf k}_+) d_{mm}^{(0)}({\bf k})<0$ of the pairing gap between electron and hole pockets. Our results for the triplet case, Eq.~\eqref{eq:LocalDOSFinal}, and the predominantly sign changing structure of d-vectors between electron and hole pockets, Fig.~\ref{fig:Dvectors}, demonstrate that the interorbital triplet pairing $\Delta(\vec L_\perp\cdot\vec \sigma_\perp)=\Delta(L_1\sigma_1+L_2\sigma_2)$ may have a qualitatively similar effect as $s_{\pm}$ singlet pairing.


%

\subsection{Spin susceptibility: Knight shift and spin resonance}
The correlation function of two operators $\hat O$ and $\hat O'$ is ($\epsilon_n^+ = \epsilon_n + \omega_m$)
\begin{eqnarray}
\chi_{OO'}({\bf q}, i \omega_m) &=& - \frac{1}{2} T \sum_{\epsilon_n} \int_{\bf k} \tr[\hat O \mathcal G(i \epsilon_n^+,{\bf k}^+)\hat O' \mathcal G(i \epsilon_n,{\bf k})] 
\end{eqnarray}
For approximately spherical Fermi surfaces $\xi_n({\bf k}) = \xi_n(k)$ and predominant intraband pairing this leads to the static spin susceptibility~\cite{MineevSamokhin1999}
 \begin{equation}\label{eq:KnightIntra}
{\chi_{S_\mu, S_\nu}^R(0,0) =   \sum_n\frac{\nu_n(E_F)}{2} \left \langle \hat d_{nn}^{(\mu)
}\hat d_{nn}^{(\nu)} Y(\hat k, T) + [\delta_{\mu \nu} - \hat d_{nn}^{(\mu)
}\hat d_{nn}^{(\nu)}] \right \rangle_{\rm FS(n)}}
\end{equation}
where $Y(\hat k, T) = \int d\xi\; {1}/({4T \cosh^2(E/2T)})$

In the limit of small $\Delta$ and no spin-orbit coupling, where Eq.~\eqref{eq:KnightIntra} is valid, the intra-band $\vec d$-vector on the Fermi surface is in the plane. This means that for superconductors with small gap the change in the spin contribution to the Knight shift will be only in the plane. On the other hand, when the gap size is comparable to the inter-band splitting, local inter-band contributions become important and the change in the Knight shift becomes purely in the $z$-direction. Therefore, the $L_1\sigma_1+L_2\sigma_2-2L_3\sigma_3$ pairing can predict different Knight shifts but it is almost always aniostropic. 

%

We now switch to the discussion of the spin resonance and the finite $\bf q$, $\Omega$ response. For purely spin singlet or purely spin triplet pairing we obtain 

\begin{eqnarray}
\chi_{S_\mu, S_\nu}^R({\bf q},\Omega)&=&  \frac{1}{4} \sum_{n,m} \sum_{hh'} \int_{\bf k} \vert \braket{{n, {\bf k}} \vert {m, {\bf k^+}}}\vert^2 \notag \\
&& \Big [ \left (\frac{\tanh\left (\frac{E_{n}({\bf k})}{2T} \right) - \tanh\left (\frac{E_{m}({\bf k^+})}{2T} \right)}{E_{n}({\bf k}) - E_{m}({\bf k^+}) + \Omega^+}  + \Omega \rightarrow - \Omega  \right )  \notag \\
&&\times \left (M_{\mu \nu}^{hh'} (u_{n, h ,{\bf k}} u_{m, h' ,{\bf k}^+} + v_{n, h ,{\bf k}} v_{m, h' ,{\bf k}^+})^2 \right )\notag \\
&& + \left (\frac{\tanh\left (\frac{E_{n}({\bf k})}{2T} \right) + \tanh\left (\frac{E_{m}({\bf k^+})}{2T} \right)}{E_{n}({\bf k}) + E_{m}({\bf k^+}) + \Omega^+}  + \Omega \rightarrow - \Omega  \right )\notag \\
&&\times \left ( M_{\mu \nu}^{hh'} (u_{n, h ,{\bf k}} v_{m, h' ,{\bf k}^+} - v_{n, h ,{\bf k}} u_{m, h' ,{\bf k}^+})^2 \right) \Big ].
\end{eqnarray}

We have introduced the matrix elements of spin-operators 
\begin{eqnarray}
M_{\mu \nu}^{hh'} &=& \frac{1}{8} \Big \lbrace [1  -hh'] (\delta_{\mu \nu} - \hat d_{(\mu} ({\bf k}) \hat d_{\nu)} ({\bf k}^+)) + [1  +hh'] \hat d_{(\mu}({\bf k}) \hat d_{\nu)}({\bf k}^+) \notag \\
&& + i \epsilon_{\mu \nu \rho} (h \hat d_\rho({\bf k}) - h'\hat d_\rho({\bf k}^+)) + hh' \delta_{\mu \nu}(1 -\hat d({\bf k}) \cdot \hat d({\bf k}^+)) \Big \rbrace,
\end{eqnarray}
and coherence factors
\begin{equation}
u_{n, h, \bk} = \frac{\xi_n + E_n}{\sqrt{2E_n(\xi_n + E_n)}}, \quad v_{n, h, \bk} = \frac{1}{\sqrt{2E_n(\xi_n + E_n)}} \times \begin{cases} d^{(0)}_{nn}, & \text{singlet,} \\ h d_{nn}, & \text{triplet.}\end{cases}
\end{equation}

The spin-resonance, as obtained by the pole of the RPA resummation of spin-interaction and bare $\chi_{S_\mu S_\nu}$ susceptibility, is most crucially determined by the last term (we omit the index $h$ in coherence factors)
\begin{eqnarray}
\sum_{hh'}M_{\mu \nu}^{hh'} (u_{n, h ,{\bf k}} v_{m, h' ,{\bf k}^+} - v_{n, h ,{\bf k}} u_{m, h' ,{\bf k}^+})^2 \stackrel{\text{singlet}}{=} \frac{\delta_{\mu \nu}}{2} (u_{n,{\bf k}} v_{m,{\bf k}^+} - v_{n,{\bf k}} u_{m, {\bf k}^+})^2.
\end{eqnarray}
In the case of singlet pairing, it approximately vanishes unless $v_{m,{\bf k}^+}v_{m,{\bf k}} <0$, i.e.~when $\Delta_{m,{\bf k}^+}\Delta_{m,{\bf k}} <0$ (we have omitted $h$ from the coherence factors since they are not $h$ dependent in the singlet case). In particular, the spin resonance is absent for $s_{++}$ pairing, while it may occur for ${\bf q}$ connecting electron and hole pockets for $s_{\pm}$.
Analogously, for the triplet case 
\begin{eqnarray}\label{eq:SpinResonance}
\sum_{hh'}M_{\mu \nu}^{hh'} (u_{n, h ,{\bf k}} v_{m, h' ,{\bf k}^+} - v_{n, h ,{\bf k}} u_{m, h' ,{\bf k}^+})^2 &\stackrel{\text{triplet}}{=}& \frac{\delta_{\mu \nu}}{2} (u_{n,{\bf k}} v_{m,{\bf k}^+} \hat d_{mm}({\bf k}^+) + v_{n,{\bf k}} u_{m, {\bf k}^+} \hat d_{nn}({\bf k}))^2 \notag \\
&&- 2 u_{n,{\bf k}} v_{m,{\bf k}^+}  v_{n,{\bf k}} u_{m, {\bf k}^+} \hat d_{nn} ^{(\nu }({\bf k}) \hat d_{mm}^{\mu )}({\bf k}^+).
\end{eqnarray}
Here, $v_{m,{\bf k}} =  v_{m,h = +,{\bf k}} $ and similarly for $u_{m,{\bf k}}$. Clearly, a sharp spin-resonance can only appear upon inclusion of SOC (and the generation of a full gap in the electronic spectrum). At the same time, the relative signs of $\vec d$-vectors, Fig.~\ref{fig:Dvectors}, demonstrates that the coherence factors [more precisely the matrix elements in Eq.~\eqref{eq:SpinResonance}] are typically non-vanishing and positive, for relative momenta connecting electron and hole pockets. This is consistent with a spin resonance at ${\bf q} = (0, \pi);(\pi, 0)$.

\subsection{Josephson scanning tunneling microscopy}

In this section we present details on Josephson scanning tunneling microscopy. 

\subsubsection{Hamiltonian of a single Josephson junction}
The Josephson Hamiltonian describing a single tunnel junction is
\begin{equation}\label{eq:JosephsonHam}
\mathcal H_{\rm Jos} = - E_C \frac{\partial^2}{\partial \phi^2} - E_J \cos(\phi - \phi_0),
\end{equation}
where $\phi$ is the current induced phase difference between tip and junction, $\phi_0$ a phase offset, $E_C$ the charging energy and the Josephson energy is $E_J = \vert I_J\vert \frac{\hbar}{e}$ ($I_J$ is the associated Josephson current), where microscopically
\begin{equation}
{
I_J({\bf x}_0) = 2\frac{e}{\hbar} \int (dE) n_F(E) \text{Im} \left \{ \tr^\sigma \left [ F_{\rm tip}^{(nn')}(E^+;{\bf x}_0, {\bf x}_0) t^{n' m}F_{\rm sample}^{(mm')}(E^+;{\bf x}_0, {\bf x}_0)  t^{m' n,*} \right ] \right \}}. \label{eq:JosephsonCurrent}
\end{equation} 
Here, Einstein summations are assumed and $ t^{m n}$ is the orbital ($m,n$) dependent tunneling element with tip position ${\bf x}_0$. 
%
%
We consider an STM tip made from  the same tRVB material as the probe. In this case, $I_J ({\bf x}_0) $ is non-zero, real, and, for a perfectly staggered order parameter, $\phi_0$ is position dependent $\phi_0({\bf x}) = \text{arg}[(-1)^{x+y}]$ 

\subsubsection{Experimental design} 

Here we design a measurement which measures $\phi_0({\bf x})$ in Eq.~\eqref{eq:JosephsonHam}. In order to ensure a coherent phase difference between sample and tip, we propose an experiment with two coherently coupled tips which form a SQUID. A simple setup is a double tip made from a single superconducting material. The staggered superconducting gap can be observed as the structure is rotated (one tip is stationary and encircled by the other), Fig.~\ref{fig:JJSpectroscopy} {\bf a}.

For an experiment which probes the phase difference between sample and tip, $E_C$ should be smaller than $E_J$. The capacitance and tunnel coupling of a superconducting tip with atomic resultion imply, however, the reverse regime $E_C \gg E_J$. Therefore, to ensure phase coherent tunneling, we propose the inclusion of a large shunt capacitor, such that $E_C \sim {e^2}/({C_{\rm tip} + C_{\rm shunt}}) \ll E_J$, see Fig.~\ref{fig:JJSpectroscopy} {\bf a}.

\subsubsection{Experimental protocol}

\begin{figure}[tp!]
\includegraphics[scale=.25]{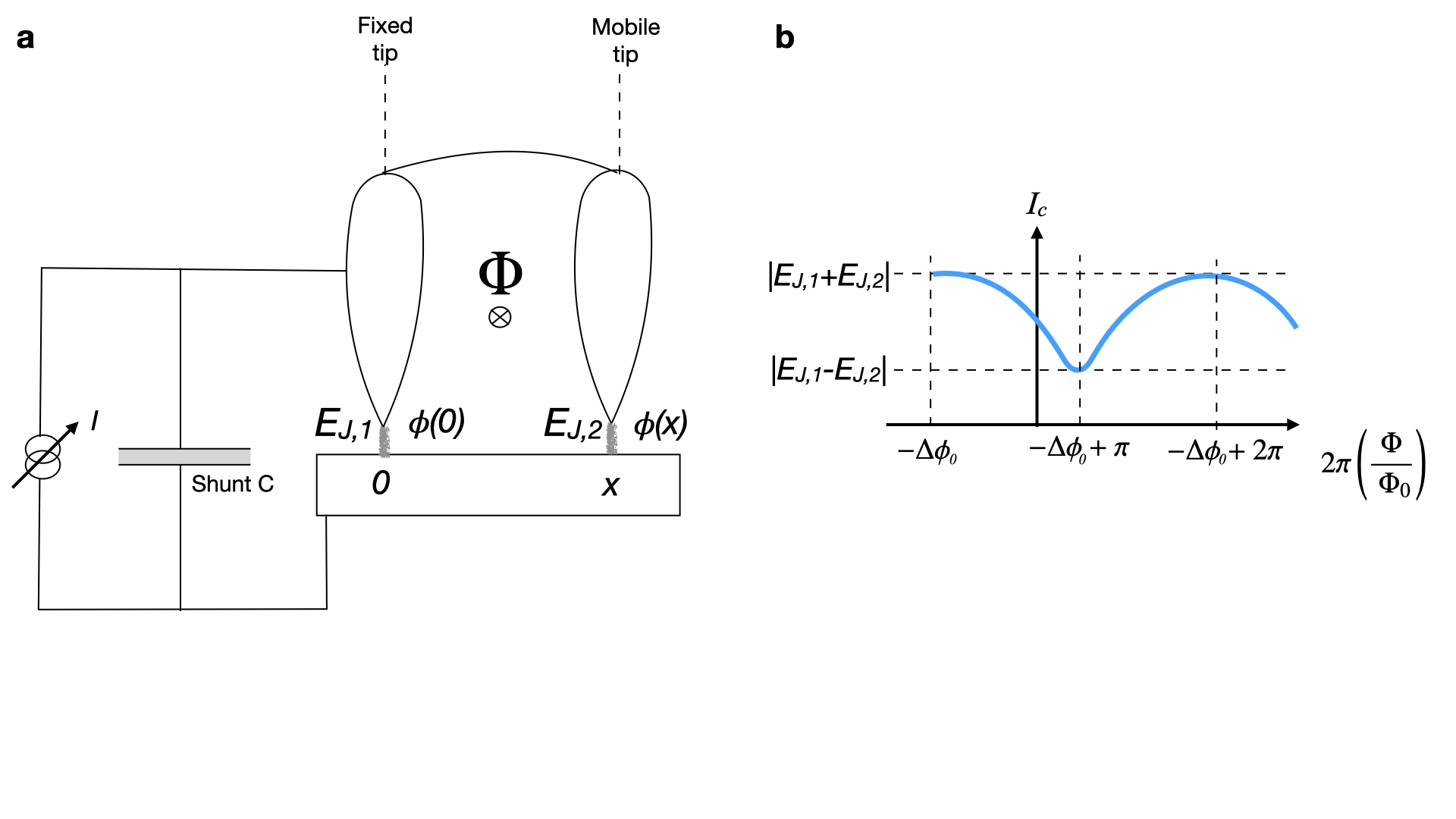}
\vspace{-2cm}
\caption{{\bf a} SQUID geometry of the proposed Josephson tunneling microscopy, including the shunt capacitor C, {\bf b} critical current as a function of applied flux $\Phi$, where $\Delta \phi_0 = \phi_0({\bf x}_1)-\phi_0({\bf x}_2)$.}
\label{fig:JJSpectroscopy}
\end{figure}

In summary, the effective Hamiltonian for the setup presented in Fig.~\ref{fig:JJSpectroscopy} {\bf a} is 
\begin{equation}
\mathcal H_{\rm SQUID} = -\text{Re} \left [e^{i\phi} \left (E_{J,1} + E_{J,2} e^{i \alpha} \right) \right],
\end{equation}
with 
$
\phi = \Delta \phi - \frac{2 e}{\hbar} A_1 L_1 + \phi_0({\bf x}_1),
\alpha = \frac{2 \pi \Phi}{\Phi_0} + \phi_0({\bf x}_2) -\phi_0({\bf x}_1)$. The critical current of the (generically asymmetric) SQUID is
\begin{equation} \label{eq:CritCurrent}
I_c = \frac{2 e}{\hbar} \sqrt{E_{J,1}^2 + E_{J,2}^2 +2E_{J,1}E_{J,2} \cos\left (\frac{2\pi \Phi}{\Phi_0} + \phi_0({\bf x}_1) - \phi_0({\bf x}_2)\right )},
\end{equation}

and presented in Fig.~\ref{fig:JJSpectroscopy} {\bf b}. 
The experimental protocol is then to determine for each pair of positions ${\bf x}_{1,2}$, whether the SQUID encloses a $\phi_0({\bf x}_1) - \phi_0({\bf x}_2)=\pi$ junction or an ordinary junction $\phi_0({\bf x}_1) - \phi_0({\bf x}_2) = 0$. This is observed by $I_c$ being minimal or maximal at zero flux. In the tRVB scenario, we expect that the phase $e^{i[\phi_0({\bf x})-\phi_0({0})]} \sim (-1)^{x+y}$ alternates between neighboring plaquettes while for ordinary superconductors no such alternation will be observed.

\end{widetext}

\bibliographystyle{apsrev4-1}
\bibliography{trvb}

\end{document}